\documentclass[comunication,amsfonts,amssymb,twocolumn,showpacs]{revtex4}
\usepackage{mathrsfs}
\usepackage{graphicx,hyperref}
\usepackage{color}
\def\be{\begin{equation}}
\def\ee{\end{equation}}
\def\ba{\begin{eqnarray}}
\def\ea{\end{eqnarray}}

\usepackage{CJK}
\begin{document}

\title{Switching management by adiabatic passage in two periodically modulated
nonlinear waveguides}

\author{Baiyuan Yang$^{1}$}

\author{Xiaobing Luo$^{1}$}
\altaffiliation{Corresponding author: xiaobingluo2013@aliyun.com}

\author{Xiaoguang Yu$^{1}$}

\affiliation{$^{1}$Department of Physics, Jinggangshan University,
Ji'an 343009, China}

\begin{abstract}
We theoretically investigate light propagation in two periodically modulated
nonlinear waveguides with certain propagation constant detuning between two guides. By slowly varying the amplitude of modulation, we can steer the light to the desired output waveguide when equal amounts of lights are launched into each waveguide. We also reveal that the light propagation dynamics depends sensitively on the detuning between two guides.
Our findings can be explained qualitatively by means of adiabatic navigation of the extended nonlinear Floquet states.

\end{abstract}

\maketitle

In recent years, a lot of interests have been made to the theoretical and experimental advances in the field of engineered photonic structures\cite{Longhi,Garanovich}. Two illuminating examples are periodically modulated waveguide arrays and directional couplers. They not only provide an ideal platform for investigating a wide variety of coherent quantum effects including coherent enhancement and destruction of tunneling\cite{Vorobeichik,Luo,Szameit1}, Zener tunneling\cite{Ghulinyan,Trompeter} and dynamical localization\cite{Longhi2}, but also open up exceptional opportunities
for the control of light propagation such as discrete diffraction-managed solitons\cite{Eisenberg,Ablowitz,Szameit}, all-optical switching of polychromatic or monochromatic light\cite{Garanovich2,Luo3}, soliton switching\cite{Papagiannis}, and so on. In addition to periodic modulation of a photonic lattice by periodically curving waveguide or varying its refractive index along the propagation direction, adiabatic passage scheme in optical waveguide system by slowly varying its geometry or refractive index is also an attractive alternative for control of light tunneling  and demonstration of adiabatic light transfer such as Landau-Zener tunneling\cite{Dreisow}, stimulated
Raman adiabatic passage in linear\cite{Longhi3} and nonlinear regimes\cite{Lahini}, and autoresonant dynamics\cite{Barak}.

Combination of periodic modulation and adiabatic management
of the system parameters provides an additional possibility for control of light propagation.
Recently, some proposals have been suggested independently for transition of a superfluid to a Mott-insulator\cite{Eckardt,Creffield,Zenesini} and generation of coherent matter currents\cite{Creffield2} in many-body systems of driven Bose-Einstein condensates (BECs), and for realization of wave packet dichotomy\cite{Longhi4} and adiabatic quantum state transfer\cite{Longhi5} in modulated linear waveguide systems, both by slowly tuning the amplitude of modulation in these periodically driven systems. In two subsequent works, the mean-field dynamics of driven BECs has been investigated by extending the conventional Floquet states of linear systems to non-linear Floquet states, and it is found that atomic population can be precisely manipulated by adiabatically controlling nonlinear Floquet state on condition that the nonlinear strength is slowly changed\cite{Gong,Molina2}. In view of the analogy between the mean-field dynamics of BEC and optics of Kerr media, such two methods for adiabatic control of nonlinear Floquet state proposed in Refs.~\cite{Gong,Molina2} may be applied to the modulated nonlinear waveguide systems. However, adiabatic management of Kerr nonlinearity is not so accessible as management of other system parameters (for example, the linear refractive index profile) in optical waveguide systems.

In this article, we consider light propagation in two periodically modulated
nonlinear waveguides with certain propagation constant detuning between two guides. We find that through adiabatical increase of the amplitude of modulation, the light becomes concentrated in a single waveguide when equal amounts of lights are launched into each waveguide, and that the final light intensity distribution is highly determined by
 the detuning between two waveguides. Our results can offer
benefits for all-optical switching and navigation of nonlinear Floquet state in the nonlinear waveguide systems.

We consider the simplest possible
arrangement which consists of two coupled asymmetric waveguide
elements with Kerr nonlinearity and with the linear refractive index periodically modulated along the propagation direction. We also suppose that each of the waveguides is single moded and excitation of radiation modes is neglected. Under these conditions, and with the use of coupled-mode theory, the evolution of the electric
field for the two-channel coupler is described by the following set of
equations:
\begin{eqnarray}
i\frac{d c_1}{dz}&=&\frac{E_0}{2}c_1+\frac{E(z)}{2}c_1-\chi|c_1|^{2} c_1-\frac{v}{2}c_2,
\label{eq1}\\
i\frac{d c_2}{dz}&=&-\frac{E_0}{2}c_2-\frac{E(z)}{2}c_2-\chi|c_2|^{2}
c_2-\frac{v}{2}c_1\label{eq2},
\end{eqnarray}
where $c_{1}$ and $c_{2}$  represent respectively field amplitudes in the first waveguide
 and the second waveguides, $\chi$ is the strength of Kerr nonlinearity, $v=\pi/L_c$ is the
coupling constant with coupling length $L_c$, $E_0$ denotes the detuning between two waveguides, and $z$ represents a
dimensionless propagation distance.  Here we
take the form of $E(z)=E_1\cos(\omega z)$
 with $E_1$ being the amplitude and
$\omega$ frequency of the modulation. It is clear that if we view $z$ as
time $t$, the above equations can be regarded as describing the
system of a quantum wave under periodic driving.

\begin{figure}[htp]
\center
\includegraphics[width=8cm]{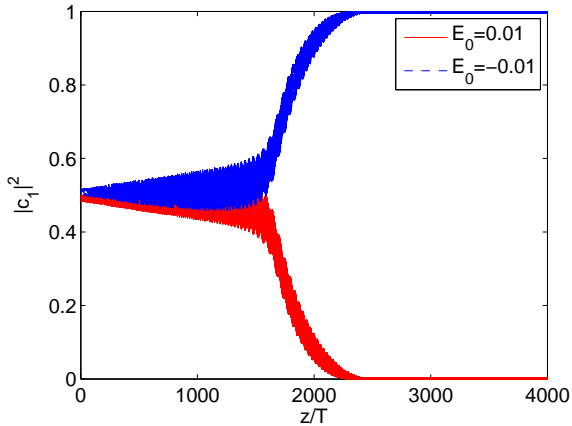}
\caption{(color online) Light localization induced from a linearly ramped
modulation $E_1(z)=Kz$, where $K=0.01T^{-1}$ with $T=2\pi/\omega$ being the modulation period, for the coupled-mode equation
(\ref{eq1}) and (\ref{eq2}) with two
different values of $E_0$.  The initial condition is $\{c_1=1/\sqrt{2},
c_2=1/\sqrt{2}\}$. The other parameters are $\chi=-0.4, v=1, \omega=10$. After
the working amplitude $E_1/\omega=2.4$ has been reached, holding $E_1(z)$ constant keeps the light localization at a
constant level.} \label{fig1}
\end{figure}

We begin by considering the light tunneling behavior when the amplitude $E_1(z)$ of the modulation
is adiabatically increased from zero to a constant value. For simplicity, we consider
a linear ramp $E_1(z)=Kz$.
The working amplitude $E_1/\omega$ is held
constant at $2.4$ after it reaches the point. We have solved the two coupled
equations (\ref{eq1}) and (\ref{eq2}) numerically with the initial state $\{c_1=1/\sqrt{2},c_2=1/\sqrt{2}\}$. Two different scenarios of the beam dynamics are identified in Fig.~(\ref{fig1}), for two different values of the detuning $E_0$. For small detuning $E_0=0.01$, we see that the light is finally localized in the second waveguide. As the detuning is changed to $E_0=-0.01$, the light becomes concentrated in the first waveguide. The numerical results clearly indicate that we can steer the light to the desired output waveguide in an adiabatical manner in the periodically modulated nonlinear couplers.
Likewise, under the circumstance that the amplitude $E_1(z)$ of the modulation
is adiabatically decreased from $E_1/\omega=2.4$ to zero, when light is launched into one waveguide, it will equally split into the two output waveguides in a reverse process.

\begin{figure}[htp]
\center
\includegraphics[width=8cm]{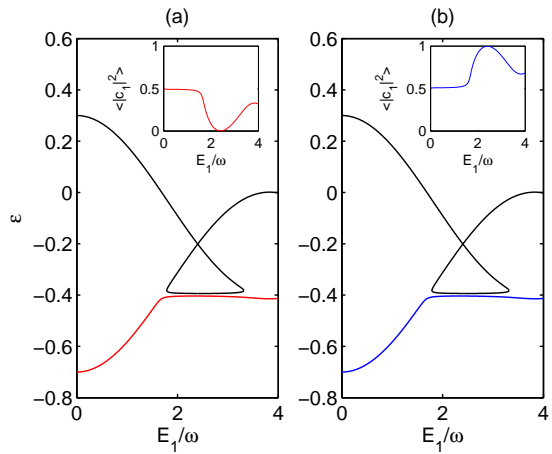}
\caption{(color online) The quasi-energies versus $E_1/\omega$ at (a) $E_0=0.01$ and (b)
$E_0=-0.01$.
The top-right inset is the time-averaged intensity $\langle |c_1|^2 \rangle$
for the Floquet state in the lowest quasi-energy level.
The other parameters are $\chi=-0.4, v=1, \omega=10$.} \label{fig2}
\end{figure}

To shed light on the underlying physics, we
turn to the Floquet theory for a periodically-driving system. Though
our system is nonlinear, its Floquet
state and quasienergy can be similarly defined. That is, Eqs.~(\ref{eq1}) and (\ref{eq2}) possess Floquet
states in the form of $(c_1,c_2)^T=(\tilde{c}_1,\tilde{c}_2)^T\exp(-i\varepsilon
z)$, where  $\varepsilon$ is  the quasi-energy and amplitudes
$(\tilde{c}_1,\tilde{c}_2)^T$ are periodic with modulation period
$T = 2\pi/\omega$.

Adopting the numerical method developed in Refs.~\cite{Luo} and \cite{Luo2}, we have computed
the Floquet states and corresponding quasienergies
$\varepsilon$. The results are plotted in Fig.~(\ref{fig2}), which shows that two extra quasienergy levels will emerge within a certain range of $E_1/\omega$ for both cases of $E_0=\pm0.01$, in stark contrast to the linear case where the number of
quasienergy levels is fixed by the size of the chosen basis. Different from the zero-detuning case
[i.e., $E_0=0$ in Eqs.~(\ref{eq1}) and (\ref{eq2})], where nonlinear Floquet
states display degeneracy in the lowest quasi-energy level at the bottom of the triangular structure\cite{Luo,Luo2}, degeneracies are lifted for a nonzero detuning $E_0\neq 0$.
We also display in the inset figures the time-averaged
population probability $\langle |c_1|^2\rangle=(\int_{0}^{T}dz
|c_1|^2)/T$ for the Floquet state corresponding
to the lowest quasienergy.
The insets show that the lowest Floquet state with nearly symmetric population distribution can undergo a strictly continuous evolution to a state with strong population imbalance. It is interesting to note that
the population imbalance of the lowest Floquet state for detuning $E_0=0.01$ is almost the opposite to that for detuning $E_0=-0.01$.
Thus, by choosing different $E_0$ signs, we can realize the strong localization of light intensity in different waveguides through adiabatic navigation of the lowest nonlinear Floquet states when the modulation amplitude $E_1$ is increased slowly from zero to a constant value [see Fig.~(\ref{fig1})].

\begin{figure}[htp]
\center
\includegraphics[width=8cm]{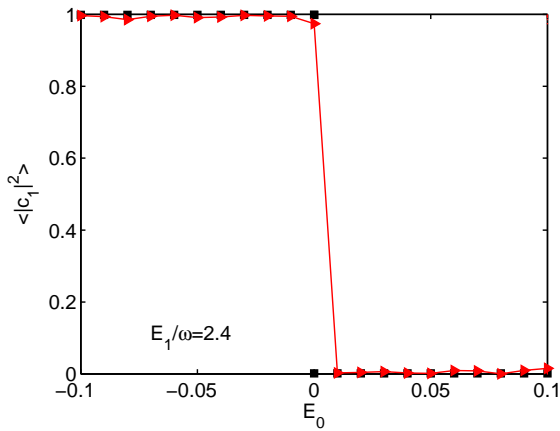}
\caption{(color online) The time-averaged intensity (black squares) for every Floquet state in the
lowest quasienergy level at the working modulation amplitude $E_1/\omega=2.4$. The red triangles are for the recorded quantities $|c_1|^2$ at $E_1/\omega=2.4$ according to the adiabatic process outlined in Fig.~(\ref{fig1}) (more details can be seen in text). The other parameters are the same as the ones in Figs.~(\ref{fig1}) and (\ref{fig2}).} \label{fig3}
\end{figure}

The induced localization versus detuning $E_0$ is more clearly demonstrated in Fig.~(\ref{fig3}). The figure shows the time-averaged intensity $\langle |c_1|^2 \rangle$ (black squares) for the lowest Floquet state at the working amplitude $E_1/\omega=2.4$ with a scan of the detuning $E_0$ across zero, which indicates that states with opposite population
imbalances can be reached through choosing different detuning signs. To describe the dynamical process, by choosing the initial state $(c_1, c_2)=(1/\sqrt{2},1/\sqrt{2})$ which is close to the ground state of the undriven system and by slowly increasing the modulation amplitude $E_1/\omega$ from zero to the working amplitude $E_1/\omega=2.4$, we record quantity $|c_1|^2$ at the end of process, illustrated as red triangles in Fig.~(\ref{fig3}). In this process, the system will adiabatically follow the lowest Floquet state and thus achieve the targeted state with complete light localization in a desired waveguide when positive or negative detuning $E_0$ is chosen. Our numerical results show that the light localization persists for moderate values of detuning $E_0$, which implies it is
easier to realize experimentally the light switching managements.

To elucidate a more rigorous dynamic, we conducted
simulations with the nonlinear Schr\"{o}dinger equation for
the dimensionless electric field
amplitude $\psi(x,z)$, which describes the propagation of monochromatic
light waves along $z$ direction\cite{Longhi,Garanovich}
\begin{eqnarray}
i\frac{\partial\psi}{\partial
z}=-\frac{1}{2}\frac{\partial^2\psi}{\partial x^2}-|\psi|^2\psi-pR(x,z)\psi.
 \label{eq:schopt}
\end{eqnarray}
Here $x$ and $z$ are the normalized transverse and longitudinal
coordinates, while $p$ describes the refractive
index amplitude. The refractive index distribution of
the waveguide coupler is given by
\begin{eqnarray}
R(x,z)&=&[1-\mu_0-\mu\cos\left(\omega z\right)]\exp\left[-\left(\frac{x-w_s/2}{w_x}\right)^6\right]\nonumber\\
&&+[1+\mu_0+\mu\cos\left(\omega
z\right)]\exp\left[-\left(\frac{x+w_s/2}{w_x}\right)^6\right],\label{index}\nonumber\\
\end{eqnarray}
with $w_s$ being the waveguide spacing, $w_x$ the channel width, $\mu$ the
longitudinal modulation amplitude, and $\omega$ the modulation
frequency. The super-Gaussian function $\exp(-x^6/w_x^6)$ describes the
profile of individual waveguides with widths $w_x$. The refractive index
change $\mu_0$ mainly defines the propagation constant mismatch $E_0$ in the coupled-mode equation.
\begin{figure}[htp]
\center
\includegraphics[width=8cm]{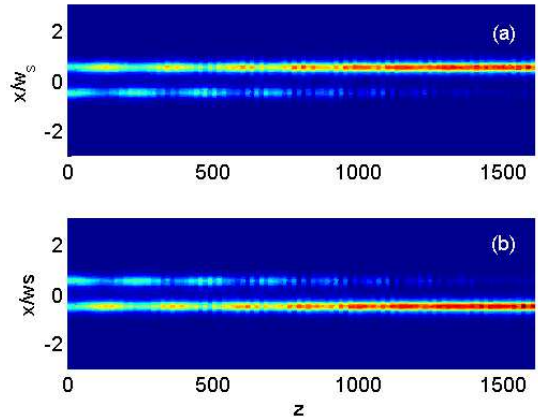}
\caption{(color online) Light propagation dynamics $|\psi(x,z)|^2$ in a modulated nonlinear coupler with the refractive index distribution Eq.~(\ref{index}) for (a) $\mu_0=-0.001$ and (b) $\mu_0=0.001$. Here the amplitude $\mu$ of the modulation
is adiabatically increased from zero to a constant value $\mu=0.2$ which corresponds to the working amplitude $E_1/\omega=2.4$ in the coupled-mode equations. The propagation distance is $L=16T_b$, and the input beam is $\psi(x,0)=0.3\phi_g(x)$ where $\phi_g(x)$ is the linear fundamental mode for the coupler with the refractive index distribution Eq.~(\ref{index}) ($\mu_0 =\mu =0$). Other parameters are given in the text.} \label{fig4}
\end{figure}

We numerically simulate the
modulated waveguide coupler by integrating the continuous wave
equation (\ref{eq:schopt}). In our simulation, the initial
states are chosen as $\psi(x,0)=A\phi_g(x)$ with $\phi_g(x)$
being the shape of the fundamental linear mode of the unmodulated symmetric coupler and $A$ the input amplitude,
and the dimensionless parameters are set as
$w_x=0.3$, $w_s=3.2$, $p=2.78$ and $\omega=3.45\times(2\pi/100)$. As
in the current experimental setup~\cite{Szameit1,Kartashov}, $w_x$ and
$w_s$ are in units of 10 $\mu$m, and $p = 2.78$ corresponds to a
refractive index of $3.1\times 10^{-4}$. When $\mu_0 =\mu =0$, the light periodically
switches between channels with beating frequency $\Omega_b=2\pi/T_b$, where $T_b=100$ for those parameters. The amplitude $\mu$ of the modulation
is adiabatically increased from zero to a constant value $\mu=0.2$ for systems
sizes up to $z=16T_b$. With the given system parameters, we firstly use the imaginary
time evolution method to find the lowest state $\phi_g(x)$ for the symmetric linear coupler ($\mu_0 =\mu =0$ in Eq.~(\ref{index})) which can be constructed as $\phi_g(x)=(1/\sqrt{2})[u_1(x)+u_2(x)]$, where $u_1$ and $u_2$ are the localized waves in the two individual waveguides. In
all simulations we used the input $\psi(x,0)=0.3\phi_g(x)$.

The behaviors of the light propagation are visualized in Fig.~(\ref{fig4}),
which illustrates strong light localization for slight detuning $\mu_0$.
 As the amplitude $\mu$ of the modulation
is adiabatically increased from zero to a constant value $\mu=0.2$, which corresponds to the working amplitude $E_1/\omega=2.4$ in the coupled-mode equations,
the light becomes concentrated in a single waveguide centered at $w_x/2$ when $\mu_0=-0.001$,
 whereas it is finally confined in the other waveguide centered at $-w_x/2$ if the detuning is changed to $\mu_0=0.001$.
 The numerical result shows that strong confinement of light in a single waveguide with relatively higher static refractive index can be achieved by slowly increasing the amplitude $\mu$ of the modulation. It is in good agreement
with the predictions based on the coupled-mode theory.

In summary, we have suggested a method for controlling light propagation in a periodically modulated nonlinear coupler by adiabatically varying the amplitude of modulation instead of varying the nonlinear strength proposed in some previous works. We find that induced light localization depends sensitively on the sign of detuning between two waveguides and thus it is possible to control the distribution of light among the output guides. The findings may offer a great potential for all-optical beam shaping and switching.

The work was supported by the NSF of China under Grants 11465009, 10965001, 11165009, the Program for New Century Excellent Talents in University of
Ministry of Education of China (NCET-13-0836), the Jiangxi Young Scientists
Training Plan under Grant No 20112BCB23024, Atomic and Molecular
Physics Key Discipline of Jiangxi Province.


\end{document}